\documentclass[11pt]{article}
\usepackage{amsmath}
\usepackage{graphicx}
\usepackage{indentfirst}
\usepackage{amssymb}
\usepackage{cite}
\usepackage{color}
\usepackage{xcolor}
\usepackage[breaklinks=true]{hyperref}
\usepackage{subfigure}
\usepackage{varwidth}
\usepackage{orcidlink}

\begin{document}
\title{Holographic Einstein Rings of AdS Black Holes in Horndeski Theory}
\date{}
\maketitle
\begin{center}
	\author{Zhi Luo\orcidlink{0009-0006-8622-2078}$~^{a,}$\footnote{zhiluo@cqu.edu.cn},
		Ke-Jian He$~^{a,c,}$\footnote{kjhe94@163.com},
		Jin Li\orcidlink{0000-0001-8538-3714}~$^{a,b,}$\footnote{cqstarv@hotmail.com (corresponding author)}}
\end{center}
\begin{center}
	$^a$ Department of Physics, Chongqing University, Chongqing 401331, China\\
	$^b$  Department of Physics and Chongqing Key Laboratory for Strongly Coupled Physics, Chongqing University, Chongqing 401331, China\\
	$^c$ Department of Mechanics, Chongqing Jiaotong University, Chongqing, 400074, China
\end{center}
\noindent
{\bf Abstract:}
By utilizing the AdS/CFT correspondence and wave optics techniques, we conducted an extensive study of the imaging properties of holographic Einstein rings in the context of Anti-de Sitter (AdS) black holes (BHs) in Horndeski theory. Our results indicate that the optical characteristics of these holographic Einstein rings are significantly influenced by the observer's position, the physical parameters of the BH, the nature of the wave source, and the configuration of the optical system. Specifically, when the observer is positioned at the north pole of the AdS boundary, the holographic image prominently displays a ring structure aligning with the BH's photon sphere. We thoroughly analyzed how various physical parameters---including the observation position, event horizon radius, temperature, and the parameter $\gamma$ in Horndeski theory---affect the holographic Einstein rings. These parameters play a crucial role in determining the rings' radius and brightness, with variations potentially causing the ring structures to deform or even transform into bright spots. Furthermore, our comparative analysis between wave optics and geometric optics reveals a strong agreement in predicting the positions and brightnesses of both the photon ring and the Einstein ring. This research offers new insights into the spacetime geometry of BHs in Horndeski theory and proposes a promising framework for exploring the gravitational duals of strongly coupled systems.

\noindent
\thispagestyle{empty}
\newpage
\setcounter{page}{1}

\section{Introduction}
\label{sec:intro}
In a landmark discovery, the Laser Interferometer Gravitational-Wave Observatory (LIGO) detectors observed gravitational waves (GWs) emanating from the merger of two stellar-mass black holes (BHs)~\cite{LIGOScientific:2016aoc,LIGOScientific:2016vlm}, thereby providing compelling direct evidence for the existence of BHs. This pivotal achievement was soon followed by the detection of additional GW signals from mergers of diverse compact objects, greatly expanding the realm of observational astronomy~\cite{LIGOScientific:2016sjg,LIGOScientific:2017vwq,LIGOScientific:2017bnn,LIGOScientific:2017ycc,LIGOScientific:2017vox}. Complementing these findings, the Event Horizon Telescope (EHT) collaboration captured the first image of a supermassive BH at the center of the M87 galaxy~\cite{EventHorizonTelescope:2019dse,EventHorizonTelescope:2019uob,EventHorizonTelescope:2019jan,EventHorizonTelescope:2019ths,EventHorizonTelescope:2019pgp,EventHorizonTelescope:2019ggy}, marking a historic milestone as the inaugural direct observation of a BH. More recently, the EHT collaboration unveiled an image of the supermassive BH at the core of our own Milky Way galaxy, $\mathrm{Sgr} \mathrm{A}^*$~\cite{EventHorizonTelescope:2022wkp,EventHorizonTelescope:2022apq,EventHorizonTelescope:2022wok,EventHorizonTelescope:2022exc,EventHorizonTelescope:2022urf,EventHorizonTelescope:2022xqj}. These images feature a dark region encircled by a luminous ring; the dark area corresponds to the BH shadow, while the bright ring signifies the photon sphere~\cite{Gralla:2019xty,Perlick:2021aok}. The morphology and dimensions of these shadows vary depending on the BH type and the underlying gravitational theory. For example, a Schwarzschild BH yields a perfectly circular shadow~\cite{Synge:1966okc,Luminet:1979nyg}, whereas a rotating Kerr BH manifests a slightly deformed, D-shaped shadow due to its spin~\cite{Bardeen:1973tla,Hioki:2008zw,Eiroa:2017uuq}. In recent years, extensive investigations into BH shadows within various modified gravity theories have revealed that these shadows harbor rich information about spacetime geometry, offering novel perspectives for probing alternative gravitational models~\cite{Amarilla:2010zq,Amarilla:2011fx,Amarilla:2013sj,Amir:2017slq,Singh:2017vfr,Mizuno:2018lxz,Vagnozzi:2019apd,Banerjee:2019nnj,Bambi:2008jg,Bambi:2010hf,Atamurotov:2013sca,Papnoi:2014aaa,Atamurotov:2015nra,Wang:2017qhh,Guo:2018kis,Yan:2019etp,Konoplya:2019sns,Bambi:2019tjh,Allahyari:2019jqz,Vagnozzi:2020quf,Khodadi:2020jij,Vagnozzi:2022moj,Tsukamoto:2014tja,Tsukamoto:2017fxq,Hu:2020usx,Zhong:2021mty,Peng:2020wun,Hou:2022eev,Hou:2021okc,Zeng:2020vsj,Zeng:2020dco,Zeng:2021dlj,Olmo:2023lil,Asukula:2023akj,Saghafi:2022pme,Nozari:2023flq,EslamPanah:2024dfq,Jafarzade:2024knc,Hendi:2022qgi,EslamPanah:2020hoj}.

At present, research on BH shadows predominantly employs ray-tracing techniques within the framework of geometric optics. However, Hashimoto and collaborators have recently introduced a wave optics-based approach grounded in the holographic principle and the AdS/CFT correspondence~\cite{Hashimoto:2019jmw,Hashimoto:2018okj}. The AdS/CFT correspondence, a concrete realization of the holographic principle, posits an equivalence between quantum gravity theories in anti-de Sitter (AdS) space and conformal field theories (CFTs) defined on its boundary~\cite{maldacena1999AdSCFT}. A prominent example of this duality is the correspondence between type IIB string theory on $AdS_5 \times S^5$ and $\mathcal{N}=4$ supersymmetric Yang-Mills theory~\cite{aharony2000large,natsuume2015ads}. This duality has been extensively applied to the study of strongly coupled systems, offering substantial theoretical insights in fields such as quantum chromodynamics (QCD) and condensed matter physics, and has further stimulated exploration into other holographic correspondences~\cite{erlich2005QCD,hartnoll2009CMT,gubser2008breaking,hartnoll2008HS,hartnoll2008building,herzog2009holographic,strominger2001ds,bredberg2011lectures,Huang:2004ai,Li:2009zs,Sheykhi:2009zv,Micheletti:2009jy,Setare:2010wt,Huang:2010zzt,Lu:2009iv,Bai:2014poa,Aprile:2012sr,Cai:2017ihd,Kusuki:2019zsp,Akers:2019nfi,Bhattacharya:2021jrn,Karndumri:2022rlf}. More specifically, Hashimoto \textit{et al.}\ investigated a $(2+1)$-dimensional boundary CFT at finite temperature, constructing the response function of the boundary quantum field theory (QFT) by introducing a temporally periodic, localized Gaussian source as a scalar field with AdS boundary conditions~\cite{Hashimoto:2019jmw,Hashimoto:2018okj}. They then transformed this response function into a holographic image of the dual BH using wave optics, resulting in an image structure analogous to an Einstein ring. This finding suggests that within a thermal QFT environment, a holographic ``Einstein ring'' image of the dual BH can be generated, providing a direct method to test gravitational duality. Future experiments designed to observe holographic images of BHs in matter could significantly advance the study of quantum gravity. Motivated by this approach, Yang \textit{et al.}\ explored the holographic Einstein rings of charged AdS BHs, revealing that while the radius of the Einstein ring remains unaffected by the chemical potential, it is notably influenced by temperature~\cite{Liu:2022cev}. Subsequent investigations have extended these studies to holographic Einstein rings in various modified gravity backgrounds~\cite{Zeng:2023zlf,Zeng:2023tjb,Zeng:2023ihy,Hu:2023mai}. In this work, we further these explorations by examining the holographic Einstein rings of AdS BHs in Horndeski theory.

Horndeski theory, recognized as the most general scalar-tensor theory with second-order field equations, has attracted renewed interest following the introduction of the covariant Galileon model~\cite{Horndeski:1974wa,Nicolis:2008in,Deffayet:2009wt,Kobayashi:2019hrl}. To address the Ostrogradsky instability inherent in such theories, Gleyzes \textit{et al.}\ proposed the beyond Horndeski framework~\cite{Gleyzes:2014dya,Gleyzes:2014qga}, within which Babichev \textit{et al.}\ discovered static BH solutions~\cite{Babichev:2017guv}. The investigation of BHs in Horndeski theory and its extensions has garnered considerable attention in the scientific community~\cite{Rinaldi:2012vy,Cisterna:2014nua,Anabalon:2013oea,Babichev:2023psy,Charmousis:2014zaa,Minamitsuji:2013ura,Babichev:2016rlq,Feng:2015oea,Cvetic:2016bxi}. In our previous work~\cite{Luo:2024avl}, we explored photon dynamics in BH spacetimes within the Horndeski framework using ray-tracing techniques and constrained the theory's parameters based on EHT observations. Additionally, we examined the observational signatures of asymmetric thin-shell wormholes (ATWs) in Horndeski theory, identifying unique features such as ``lens bands'' and ``photon ring sets'' that distinguish BHs from wormholes~\cite{Luo:2023wru}. Building upon the methodology of Hashimoto \textit{et al.}, this paper employs wave optics to investigate the holographic Einstein rings of AdS BHs in Horndeski theory, analyzing how the holographic images correlate with the parameters of the theory and the properties of the wave sources.

The structure of this paper is as follows: In Section~\ref{sec2}, we provide an overview of AdS BHs within the Horndeski theory framework, detailing the holographic setup and the lensing response function in this spacetime. Section~\ref{sec3} discusses the utilization of an optical system comprising convex lenses and a spherical screen to study the holographic Einstein rings of AdS BHs in Horndeski theory, along with a comparative analysis of wave optics and geometric optics results. Finally, Section~\ref{secfour} concludes the paper.

\section{Response function for AdS BHs in Horndeski theory}
\label{sec2}
To initiate our analysis, we consider the action of Horndeski gravity with a negative cosmological constant $\Lambda$, expressed as~\cite{Horndeski:1974wa}:
\begin{equation}\label{eq1}
	S=\int d^4 x \sqrt{-g}\left(\mathcal{L}_2+\mathcal{L}_3+\mathcal{L}_4+\mathcal{L}_5-2\Lambda\right),
\end{equation}
with
\begin{equation}
	\mathcal{L}_2=G_2(X),
\end{equation}
\begin{equation}
	\mathcal{L}_3=-G_3(X) \square \phi,
\end{equation}
\begin{equation}
	\mathcal{L}_4=G_4(X) \mathcal{R}+G_{4 X}\left[(\square \phi)^2-\left(\nabla_\mu \nabla_\nu \phi\right)^2\right],
\end{equation}
\begin{equation}
	\mathcal{L}_5=G_5(X) G_{\mu \nu} \nabla^\mu \nabla^\nu \phi-\frac{G_{5 X}}{6}\left[(\square \phi)^3-3 \square \phi\left(\nabla_\mu \nabla_\nu \phi\right)^2+2\left(\nabla_\mu \nabla_\nu \phi\right)^3\right].
\end{equation}
In this context, $\mathcal{R}$ denotes the Ricci scalar, and $G_{\mu\nu}$ represents the Einstein tensor. The scalar field is $\phi$, with its kinetic term defined as $X = -\frac{1}{2} \partial_\mu \phi \partial^\mu \phi$. The functions $G_i = G_i(\phi, X)$ depend on both $\phi$ and $X$, where the subscript $X$ indicates differentiation with respect to $X$. Building upon the work in Ref.~\cite{Babichev:2017guv}, we consider a specific case of the Horndeski action~(\ref{eq1}), characterized by:
\begin{equation}
	G_2 = \eta X, \quad G_4 = \zeta + \beta \sqrt{-X}, \quad G_3 = G_5 = 0.
\end{equation}
By setting $\zeta = \frac{1}{2}$ and considering $\eta$ and $\beta$ as dimensionless parameters, the action in Eq.~(\ref{eq1}) reduces to
\begin{equation}\label{eqac}
	\begin{aligned}
		S= & \int \mathrm{d}^4 x \sqrt{-g}\left\{\left[\frac1{2}+\beta \sqrt{(\partial \phi)^2 / 2}\right] \mathcal{R}-\frac{\eta}{2}(\partial \phi)^2-\frac{\beta}{\sqrt{2(\partial \phi)^2}}\left[(\square \phi)^2-\left(\nabla_\mu \nabla_\nu \phi\right)^2\right]-2\Lambda\right\}.
	\end{aligned}
\end{equation}
A spherically symmetric AdS BH solution in Horndeski theory can be derived from Eq.~(\ref{eqac}) and is given by~\cite{Babichev:2017guv}:  
\begin{equation}\label{eqdsff}
	\mathrm{d} s^2=-f(r) \mathrm{d} t^2+\frac{1}{f(r)} \mathrm{d} r^2+r^2\left(\mathrm{d} \theta^2+\sin ^2 \theta \mathrm{d} \varphi^2\right),
\end{equation}
in which
\begin{equation}\label{eqfgamma}
	f(r)=1-\frac{2M}{r}-\frac{\gamma}{r^2}+\frac{r^2 }{l^2}.
\end{equation}
To simplify the expressions in Eq.~(\ref{eqfgamma}), we introduce the parameter $\gamma = \beta^2 / \eta$. Here, $l = \sqrt{-3\Lambda}$ denotes the radius of the AdS spacetime, and the constant $M$ acts as an integration constant corresponding to the mass of the BH. By setting $f(r_h) = 0$, we determine the radius $r_h$ of the BH's outer event horizon. Consequently, the BH mass can be re-expressed in terms of the event horizon radius $r_h$ as
\begin{equation}\label{unitl}
	M = \frac{r_h^2 + r_h^4 - \gamma}{2 r_h},
\end{equation}
where we have adopted the unit $l = 1$ in Eq.~(\ref{unitl}) for simplicity.

To develop a holographic model for AdS BHs in Horndeski theory, we reformulate the metric function by introducing a new variable $\xi = 1/r$. This substitution leads to the following expression:
\begin{equation}\label{frxi}
	f(r) = \frac{1}{\xi^2} f(\xi).
\end{equation}
Furthermore, the Hawking temperature of the BH is
\begin{equation}
	T = \frac{3+\xi_h^2+\xi_h^4 \gamma}{4 \pi \xi_h},
\end{equation}
where $\xi_h$ is the event horizon of the BH. Moreover, one can express the metric as follows:
\begin{align}
	ds^2 = \frac{1}{\xi^2}\left[ -f(\xi)\, dt^2 + \frac{1}{f(\xi)}\, d\xi^2 + d\theta^2 + \sin^2\theta\, d\varphi^2 \right], \label{metric3}
\end{align}
where $f(\xi)$ is the metric function redefined in terms of $\xi$. To simplify calculations near the event horizon and to analyze wave propagation more effectively, we adopt the ingoing Eddington–Finkelstein coordinates $(v, \xi, \theta, \varphi)$, defined by
\begin{equation}
	v = t + \xi_* = t - \int \frac{d\xi}{f(\xi)}.
\end{equation}
In these new coordinates, the metric \eqref{metric3} transforms to
\begin{align}\label{eqKG1}
	ds^2 = \frac{1}{\xi^2}\left[ -f(\xi)\, dv^2 - 2\, dv\, d\xi + d\theta^2 + \sin^2\theta\, d\varphi^2 \right].
\end{align}

Now we consider an evolution of the scalar field $\Phi$, which is governed by the Klein-Gordon equation,
\begin{equation}\label{eqKG}
	\square \Phi(v, \xi, \theta, \varphi) = 0.
\end{equation}
Combining Eqs.~(\ref{eqKG1}) and (\ref{eqKG}), we arrive at
\begin{align}\label{eq11}
	\xi^2 f(\xi) \partial_\xi^2 \Phi + \left[ \xi^2 f'(\xi) - 2 \xi f(\xi) \right] \partial_\xi \Phi - 2 \xi^2 \partial_v \partial_\xi \Phi + 2 \xi \partial_v \Phi + \xi^2 D_S^2 \Phi = 0,
\end{align}
where $f'(\xi) = \partial_\xi f(\xi)$, and $D_S^2$ denotes the scalar Laplacian on the unit two-sphere $S^2$. Following Ref.~\cite{Liu:2022cev}, the asymptotic solution of Eq.~(\ref{eq11}) near the AdS boundary is given by
\begin{equation}\label{eq12}
	\Phi(v, \xi, \theta, \varphi) = \mathcal{J}_O(v, \theta, \varphi) + \xi \partial_v \mathcal{J}_O(v, \theta, \varphi) + \frac{1}{2} \xi^2 D_S^2 \mathcal{J}_O(v, \theta, \varphi) + \xi^3 \langle O \rangle + \mathcal{O}(\xi^4),
\end{equation}
where $\mathcal{J}_O(v, \theta, \varphi)$ is the scalar source term, and $\langle O \rangle$ is the corresponding response function. For a monochromatic, axisymmetric Gaussian wave packet source situated at the South Pole of the AdS boundary, we have~\cite{Zeng:2023zlf}
\begin{align}\label{eq13}
	\mathcal{J}_O(v, \theta) = e^{i \omega v} \cdot \frac{1}{2 \pi \sigma^2} \exp\left[ -\frac{(\pi - \theta)^2}{2 \sigma^2} \right] = e^{i \omega v} \sum_{l=0}^\infty C_{l0} Y_{l0}(\theta),
\end{align}
where
\begin{align}\label{eq14}
	C_{l0} = (-1)^l \sqrt{\frac{l + 1/2}{2 \pi}} \exp\left[ -\frac{1}{2} (l + 1/2)^2 \sigma^2 \right].
\end{align}
Here, $\sigma$ characterizes the width of the Gaussian wave packet, and $Y_{l0}$ are the spherical harmonics. In the limit $\sigma \ll \pi$, the scalar field $\Phi(v, \xi, \theta, \varphi)$ can be expanded as
\begin{align}\label{eq15}
	\Phi(v, \xi, \theta, \varphi) = e^{i \omega v} \sum_{l=0}^\infty c_{l0} U_l(\xi) Y_{l0}(\theta, \varphi),
\end{align}
and the response function can be expressed as
\begin{align}\label{eq16}
	\langle O \rangle = e^{i \omega v} \sum_{l=0}^\infty \langle O \rangle_l Y_{l0}(\theta).
\end{align}
Substituting Eq.~(\ref{eq15}) into the field equation, we find
\begin{align}\label{eq17}
	\xi^2 f(\xi) U''_l + \left[ \xi^2 f'(\xi) - 2 \xi f(\xi) + 2 i \omega \xi^2 \right] U'_l + \left[ -2 i \omega \xi - l(l + 1) \xi^2 \right] U_l = 0,
\end{align}
with the asymptotic behavior of $U_l$ near the boundary given by
\begin{equation}\label{eq18}
	\lim_{\xi \rightarrow 0} U_l = 1 - i \omega \xi + \frac{1}{2} \left[ -l(l + 1) \right] \xi^2 + \langle O \rangle_l \xi^3 + \mathcal{O}(\xi^4).
\end{equation}
The function $U_l$ satisfies two boundary conditions. The first is at the horizon $\xi = \xi_h$:
\begin{equation}\label{eqb1}
	\left[ \xi_h^2 f'(\xi_h) + 2 i \omega \xi_h^2 \right] U'_l - \left[ 2 i \omega \xi_h + l(l + 1) \xi_h^2 \right] U_l = 0,
\end{equation}
and the second is the AdS boundary condition at $\xi = 0$:
\begin{equation}\label{eqb2}
	U_l(0) = 1.
\end{equation}
By imposing the boundary conditions (\ref{eqb1}) and (\ref{eqb2}), the differential equation (\ref{eq17}) for $U_l$ can be solved. Subsequently, the total response function $\langle O \rangle$ can be determined using Eqs.~(\ref{eq16}) and (\ref{eq18}).
\begin{figure}[!ht]
	\centering
	\subfigure[$\gamma=-0.3$ \label{f1a}]{
		\begin{minipage}[t]{0.5\linewidth}
			\centering
			\includegraphics[width=6cm,height=6cm]{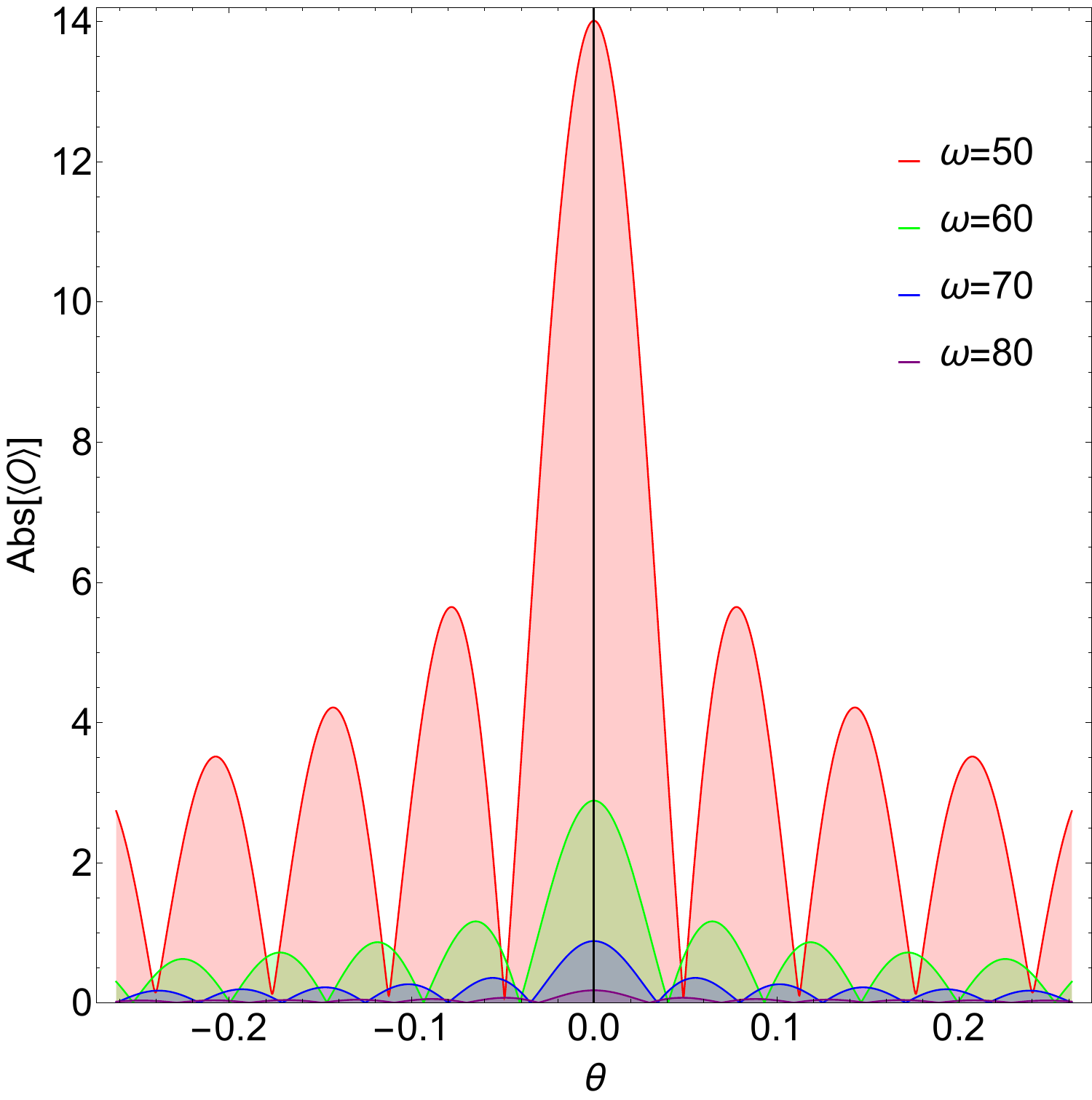}
			%\caption{The effective potential for various parameters $\gamma$}
		\end{minipage}%
	}%
	\subfigure[$\omega=80$\label{f1b}]{
		\begin{minipage}[t]{0.5\linewidth}
			\centering
			\includegraphics[width=6cm,height=6cm]{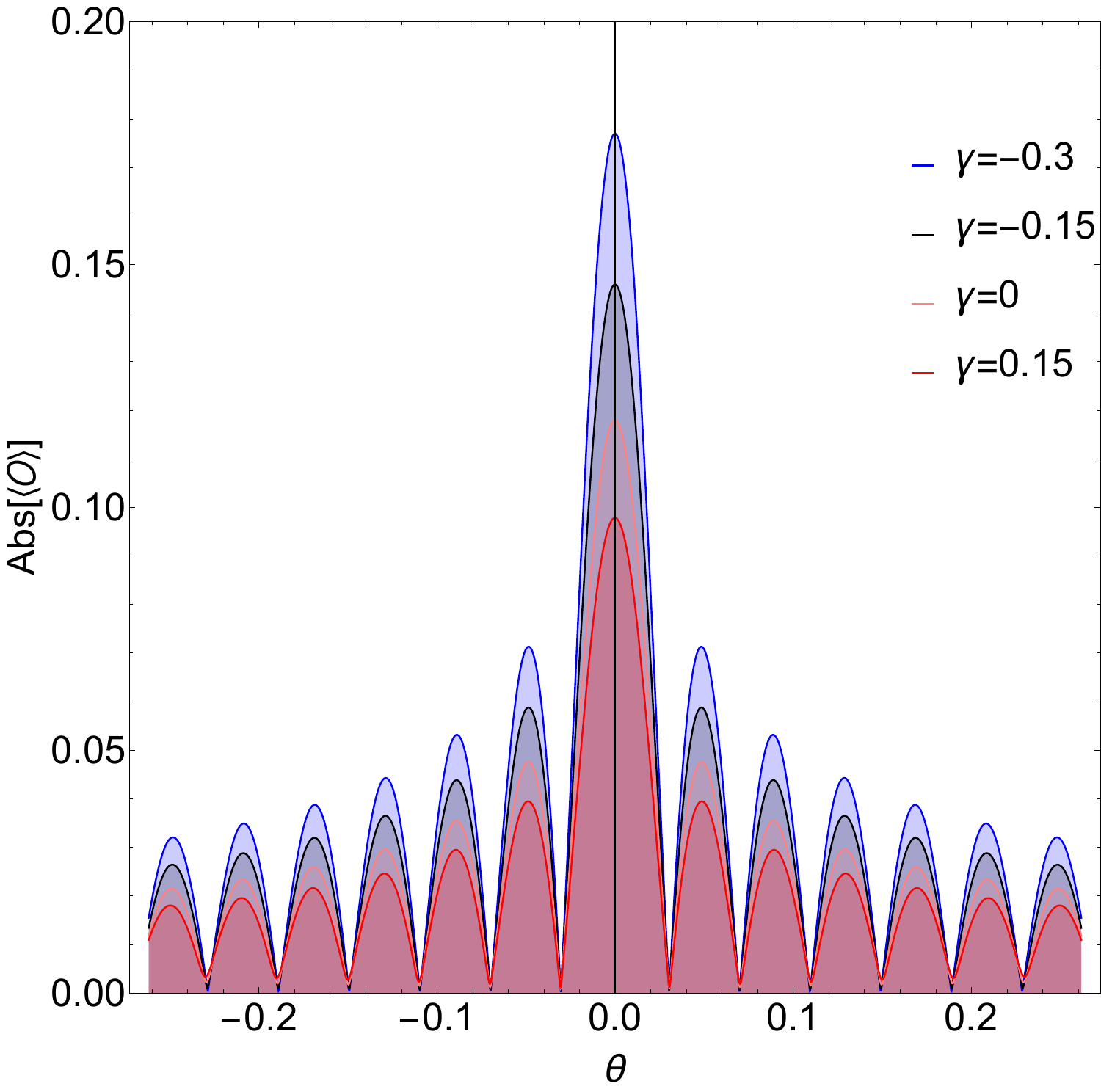}
			%\caption{The effective potential for various parameters $\gamma$}
		\end{minipage}%
	}%
	\caption{Effects of different source frequencies $\omega=50, 60, 70, 80$ and parameters $\gamma=-0.3, -0.15, 0, 0.15$ on the response function. In this analysis, we set $\xi_h = 1$.} 
	\label{f1}
\end{figure}\textbf{}

\begin{figure}[!ht]
	\centering
	\subfigure[$\gamma=-0.3$ \label{f2a}]{
		\begin{minipage}[t]{0.5\linewidth}
			\centering
			\includegraphics[width=6cm,height=6cm]{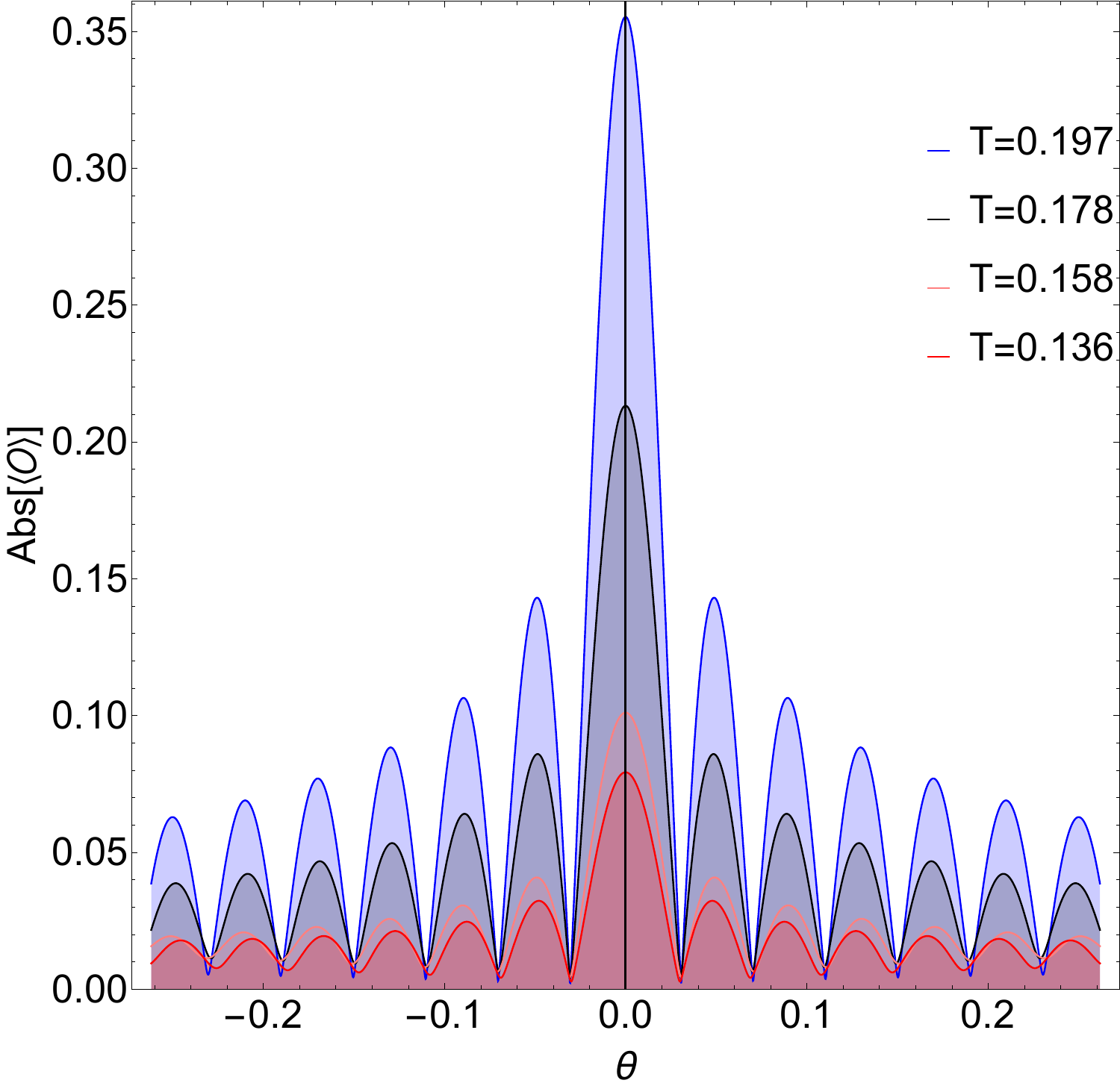}
			%\caption{The effective potential for various parameters $\gamma$}
		\end{minipage}%
	}%
	\subfigure[$\gamma=0.3$ \label{f2b}]{
		\begin{minipage}[t]{0.5\linewidth}
			\centering
			\includegraphics[width=6cm,height=6cm]{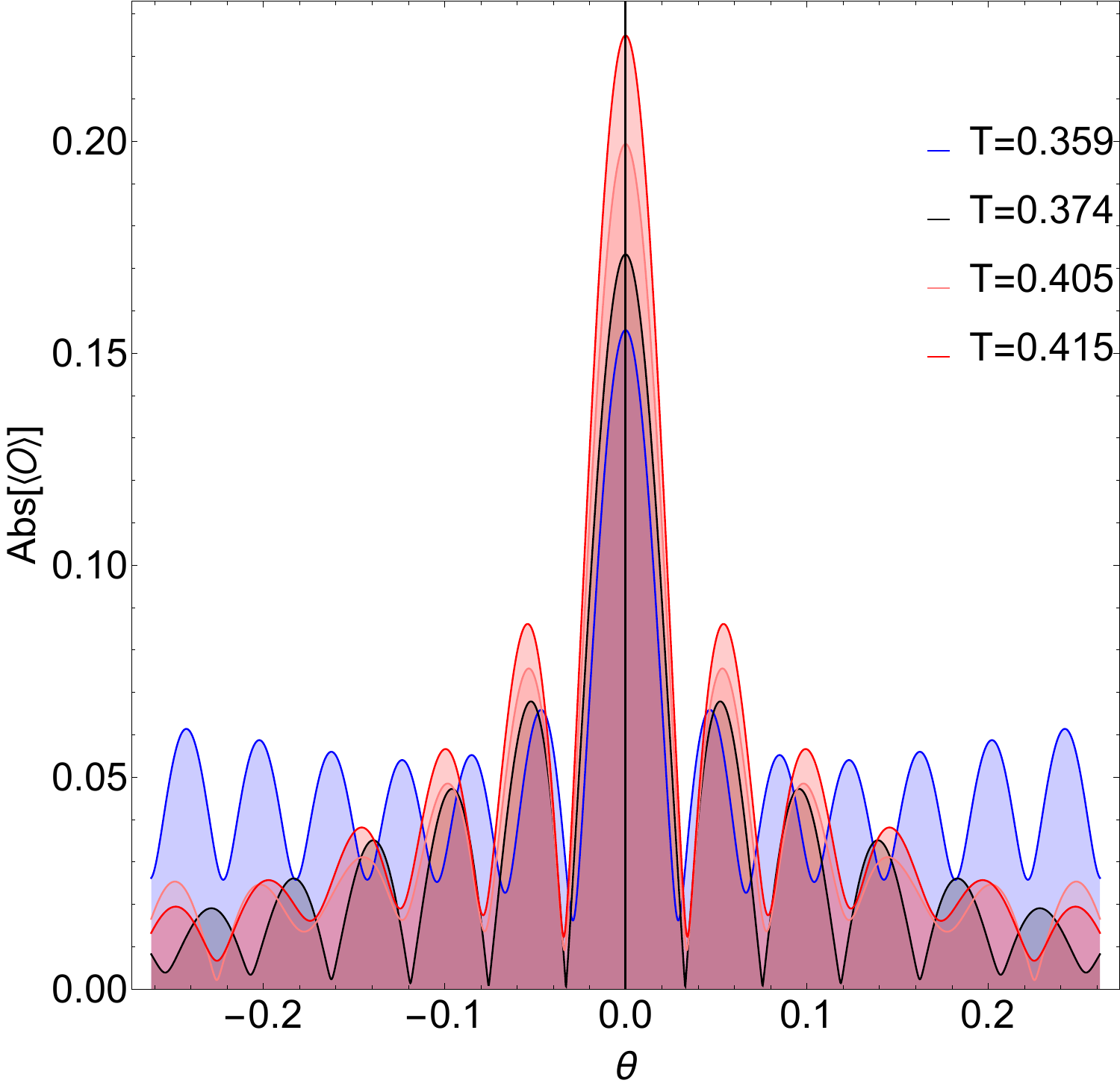}
			%\caption{The effective potential for various parameters $\gamma$}
		\end{minipage}%
	}%
	\caption{Effects of different temperatures on the response function. In this analysis, we set $\omega=80$.} 
	\label{f2}
\end{figure}\textbf{}
In Fig.~\ref{f1}, we display the amplitude of the response function as a function of source frequency $\omega$ for various values of the parameter $\gamma$. Fig.~\ref{f1a} demonstrates that as the source frequency $\omega$ increases, both the amplitude and the period of the response function decrease. In contrast, Fig.~\ref{f1b} shows that the amplitude of the response function diminishes with increasing values of the parameter $\gamma$. Furthermore, Fig.~\ref{f2} explores the relationship between the amplitude of the response function and the temperature $T$ for both positive and negative values of $\gamma$. The findings suggest that in both scenarios, the amplitude of the response function increases with rising temperature.

\section{Holographic rings of AdS BHs in Horndeski theory}
\label{sec3}
\begin{figure}[!ht]
	\centering
	\subfigure[\label{qq1}]{
		\begin{minipage}[t]{0.5\linewidth}
			\centering
			\includegraphics[width=8cm,height=4.8cm]{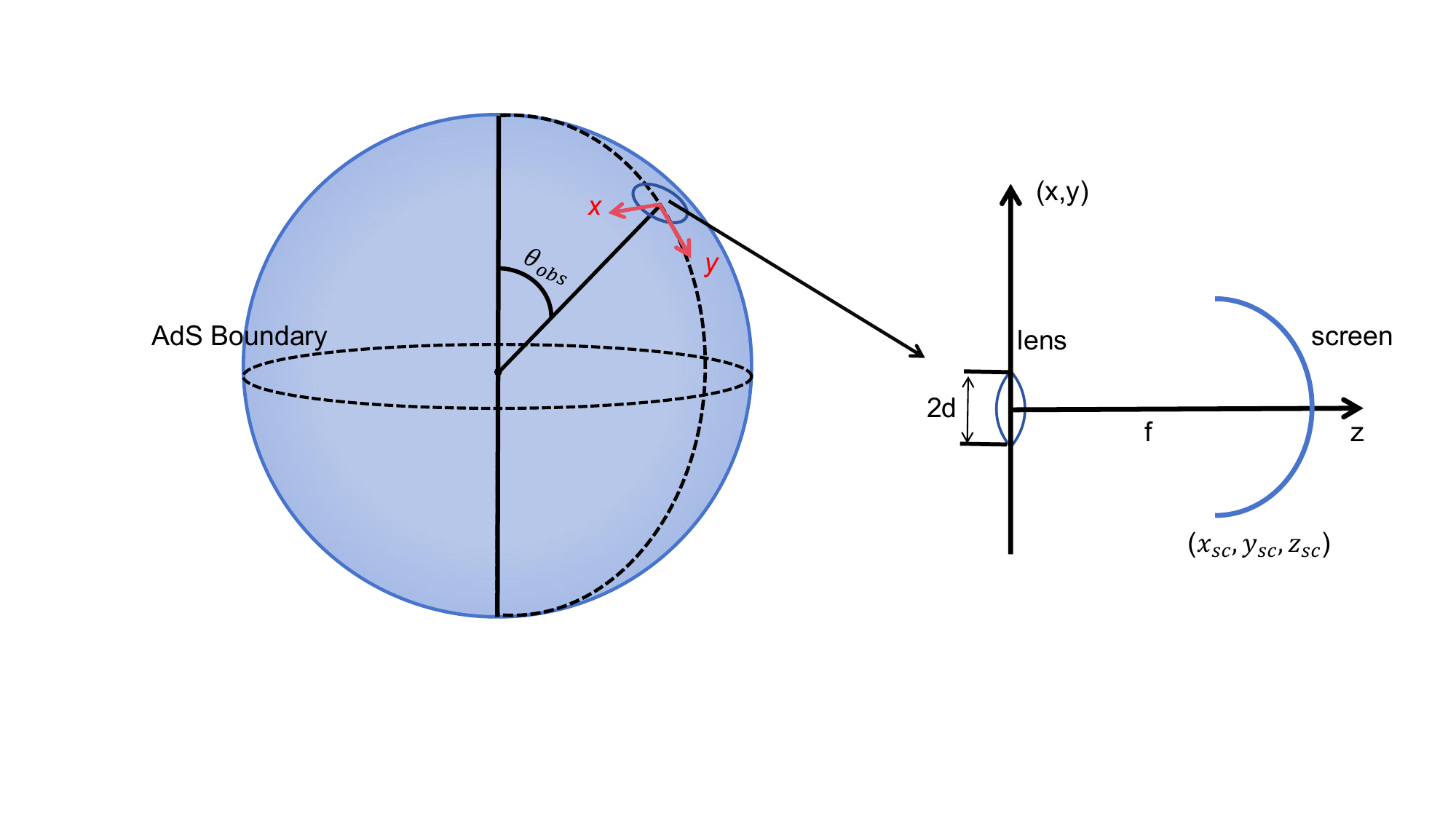}
			%\caption{The effective potential for various parameters $\gamma$}
		\end{minipage}%
	}%
	\subfigure[\label{qq4}]{
		\begin{minipage}[t]{0.5\linewidth}
			\centering
			\includegraphics[width=8cm,height=4.8cm]{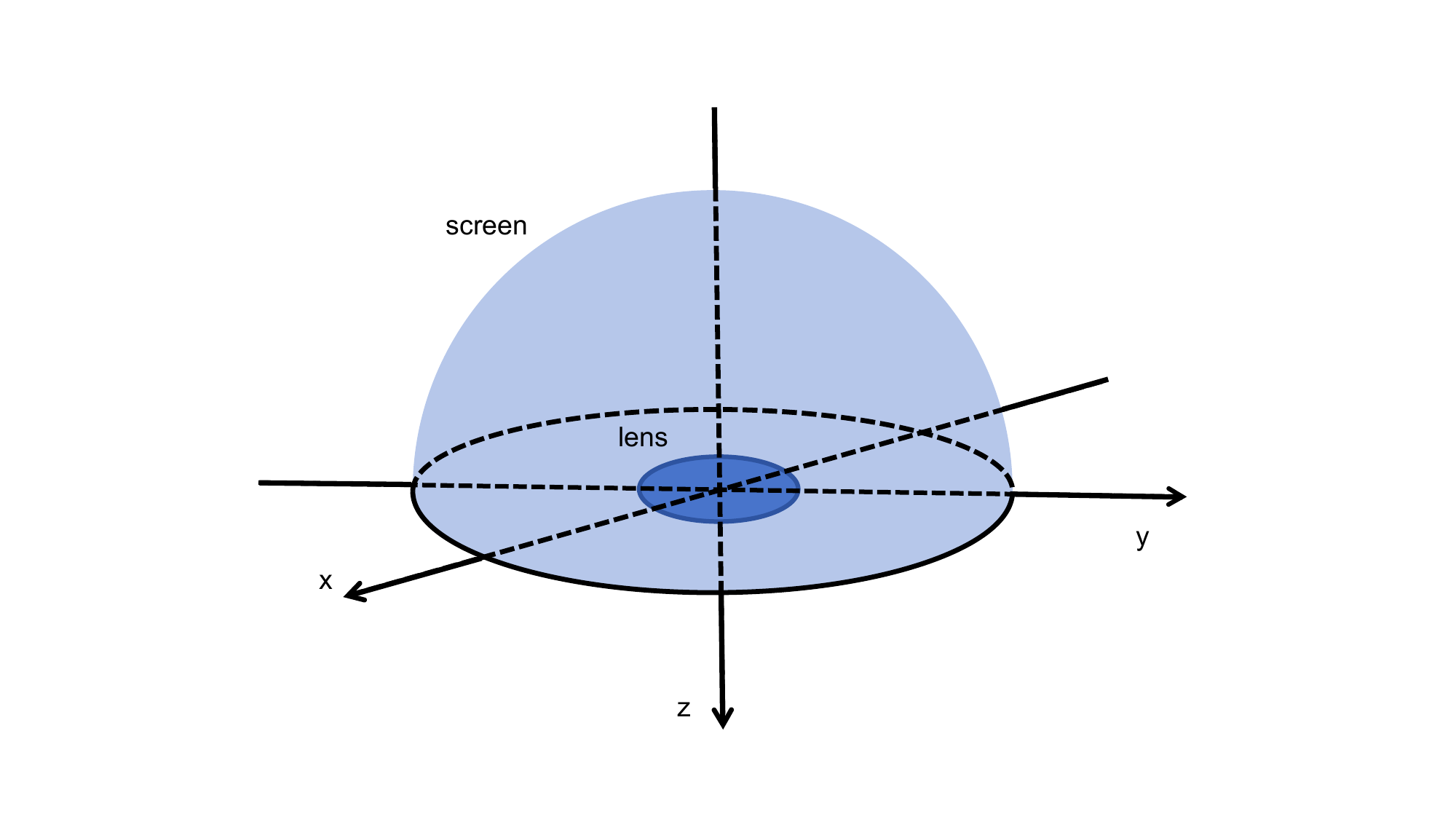}
			%\caption{The effective potential for various parameters $\gamma$}
		\end{minipage}%
	}%
	\caption{The optical system} 
	\label{figre2}
\end{figure}

In this section, we introduce the optical system shown in Fig.~\ref{figre2}(a) to simulate the imaging of the Einstein ring and subsequently explore its wave characteristics. The system consists of a spherical screen and an infinitely thin convex lens, with the focal length $f$ of the lens being much greater than its size $d$. The observation center is set at $(\theta_{obs}, 0)$, marked in blue in Fig.~\ref{figre2}(a). This optical system is positioned at the observation center, with the lens at $\vec{x} = (x, y, 0)$ and the screen at $\vec{x}_S = (x_S, y_S, z_S)$. The response function is first transformed into a transmitted wave $\Psi_s(\vec{x})$ by the convex lens, which is then projected onto the screen and converted into an observable wave $\Psi_{sc}(\vec{x}_S)$, as depicted in Fig.~\ref{figre2}(b). 
\textbf{}
Therefore, one can obtain~\cite{Zeng:2023zlf} 
\begin{equation}\label{eq19}
\Psi_{sc}(\vec{x}_S) = \int_{|\vec{x}|<d} dx^2 \Psi_s(\vec{x}) e^{- i \omega L} = \int_{|\vec{x}|<d} dx^2 e^{- i \omega \frac{|\vec{x}|^2}{2f}} \Psi_p(\vec{x}) e^{- i \omega L},
\end{equation}
in which $L$ represents the distance between $\vec{x}$ and $\vec{x}_S$, while $f$ is defined by 
\begin{equation}
	f^2 = x_S^2 + y_S^2 + z_S^2.
\end{equation}
By further combining the expression
\begin{equation}
	L = \sqrt{(x_s - x)^2 + (y_s - y)^2 + z_s^2} \simeq f - \frac{\vec{x}s \cdot \vec{x}}{f} + \frac{|\vec{x}|^2}{2f},
\end{equation}
where the Fresnel approximation $f \gg |\vec{x}|$, Eq. (\ref{eq19}) can be expressed as
\begin{equation}\label{eq21}
\Psi_{sc}(\vec{x}_S) \propto \int_{|\vec{x}|<d} dx^2 \Psi_p(\vec{x}) \varpi(\vec{x}) e^{-\frac{i \omega }{f} \vec{x}\cdot \vec{x}_s},
\end{equation}
where the window function is
\begin{align} \label{Eq22}
  { \varpi(\vec{x})}=
    \begin{cases}
   \text{1}, \quad \quad 0 \leq \mid \vec{x}\mid \leq d\\
     \text{0}, \quad \quad  \mid \vec{x}\mid > d
    \end{cases}.
\end{align}

\begin{figure}[!ht]
	\centering
	\subfigure[$\theta_{obs} = 0$, $\gamma = -0.3$ \label{b1}]{
		\begin{minipage}[t]{0.33\linewidth}
			\centering
			\includegraphics[width=4cm,height=4cm]{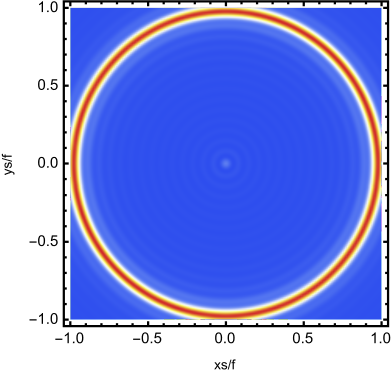}
			%\caption{fig1}
		\end{minipage}%
	}%
	\subfigure[$\theta_{obs} = \pi/4$, $\gamma = -0.3$  \label{b2}]{
		\begin{minipage}[t]{0.33\linewidth}
			\centering
			\includegraphics[width=4cm,height=4cm]{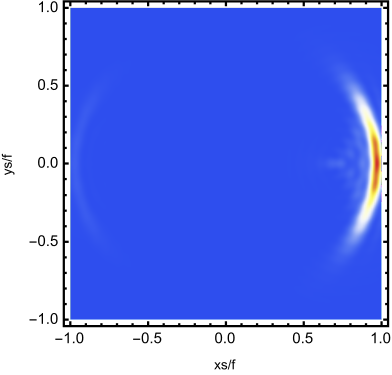}
			%\caption{fig2}
		\end{minipage}%
	}%
	\subfigure[$\theta_{obs} = \pi/2$, $\gamma = -0.3$ \label{b3}]{
		\begin{minipage}[t]{0.33\linewidth}
			\centering
			\includegraphics[width=4cm,height=4cm]{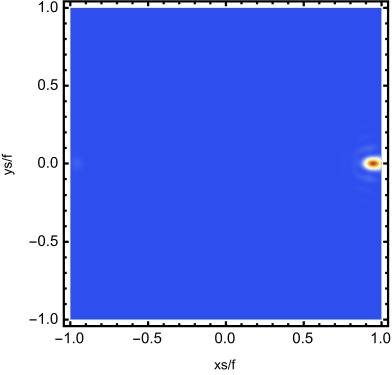}
			%\caption{fig2}
		\end{minipage}
	}%
	\quad
	\subfigure[$\theta_{obs} = 0$, $\gamma = 0$  \label{b4}]{
		\begin{minipage}[t]{0.33\linewidth}
			\centering
			\includegraphics[width=4cm,height=4cm]{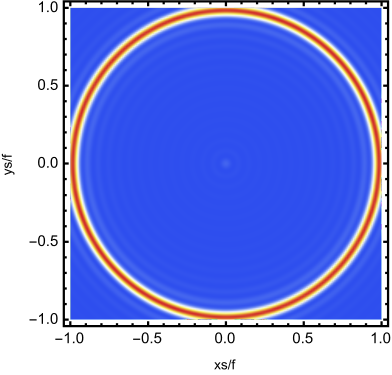}
			%\caption{fig2}
		\end{minipage}
	}%
	\subfigure[$\theta_{obs} = \pi/4$, $\gamma = 0$ \label{b5}]{
		\begin{minipage}[t]{0.33\linewidth}
			\centering
			\includegraphics[width=4cm,height=4cm]{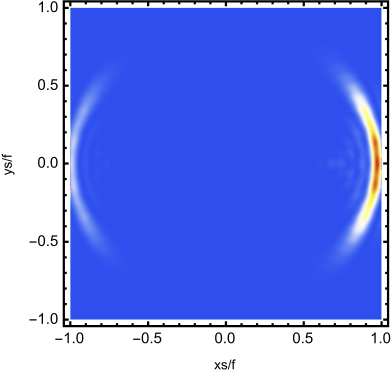}
			%\caption{fig2}
		\end{minipage}
	}%
	\subfigure[$\theta_{obs} = \pi/2$, $\gamma = 0$  \label{b6}]{
		\begin{minipage}[t]{0.33\linewidth}
			\centering
			\includegraphics[width=4cm,height=4cm]{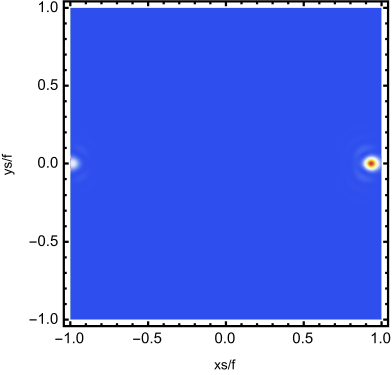}
			%\caption{fig2}
		\end{minipage}
	}%	
	\quad
	\subfigure[$\theta_{obs} = 0$, $\gamma = 0.3$  \label{b7}]{
		\begin{minipage}[t]{0.33\linewidth}
			\centering
			\includegraphics[width=4cm,height=4cm]{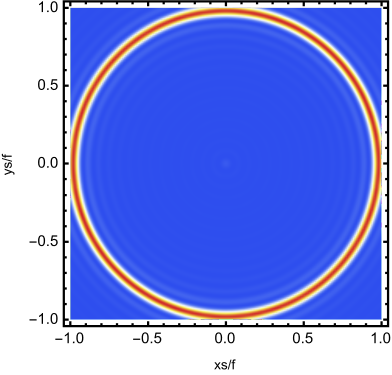}
			%\caption{fig2}
		\end{minipage}
	}%
	\subfigure[$\theta_{obs} = \pi/4$, $\gamma = 0.3$  \label{b8}]{
		\begin{minipage}[t]{0.33\linewidth}
			\centering
			\includegraphics[width=4cm,height=4cm]{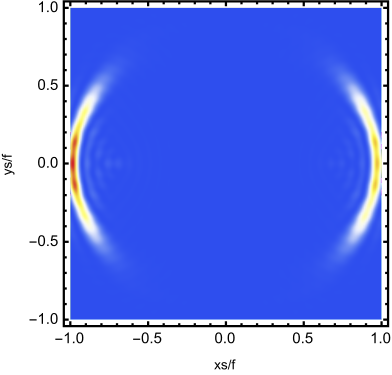}
			%\caption{fig2}
		\end{minipage}
	}%
	\subfigure[$\theta_{obs} = \pi/2$, $\gamma = 0.3$  \label{b9}]{
		\begin{minipage}[t]{0.33\linewidth}
			\centering
			\includegraphics[width=4cm,height=4cm]{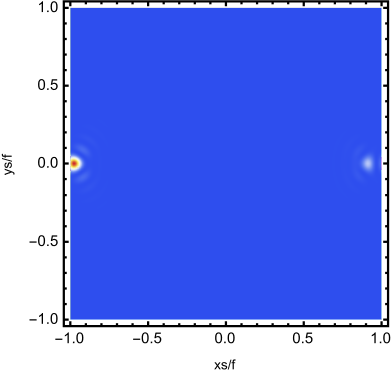}
			%\caption{fig2}
		\end{minipage}
	}%	
	\caption{Holographic images at different observation angles $\theta_{obs}$ and parameter values $\gamma$ for the AdS BH in Horndeski theory. In this analysis, we set $\xi_h = 1$ and $\omega=80$.
	}
	\label{b19}
\end{figure}
Eq.~(\ref{eq21}) demonstrate that the incident wave can be transformed into the observed wave on the screen via a Fourier transformation, thereby delineating the profiles of the double BHs on the observer's screen. In Fig.~\ref{b19}, we present holographic Einstein images at various observation angles $\theta$ and different values of the parameter $\gamma$. The results reveal that the holographic Einstein images exhibit significant variations depending on the observation angle and the value of $\gamma$. When the observation angle is $\theta = 0^\circ$, with the observer positioned at the north pole of the AdS boundary, Figs.~\ref{b1}, \ref{b4}, and \ref{b7} display the holographic Einstein images as a series of axially symmetric concentric rings, with the brightest ring located at the center, forming a Poisson-like spot. Notably, the brightness of these rings remains relatively unchanged as the parameter $\gamma$ increases. As the observation angle increases to $\theta = 45^\circ$, Figs.~\ref{b2}, \ref{b5}, and \ref{b8} show that the holographic Einstein images transition into two arcs with varying brightness, where the left arc appears dimmer than the right one. As $\gamma$ increases, the brightness of the left arc gradually intensifies. Finally, when the observation angle reaches $\theta = 90^\circ$, Figs.~\ref{b3}, \ref{b6}, and \ref{b9} indicate that the two arcs further evolve into two small bright spots. As the parameter $\gamma$ increases, the brightness of the left spot also gradually enhances. These findings underscore that both the observation angle $\theta$ and the parameter $\gamma$ have a significant influence on the morphology and brightness of the holographic Einstein images. This suggests that the spacetime characteristics of the BH are effectively encoded in these holographic images, offering a potential tool for probing the underlying geometry of BHs through observational data.

\begin{figure}[!ht]
	\centering
	\subfigure[$\omega = 20$ \label{c1}]{
		\begin{minipage}[t]{0.33\linewidth}
			\centering
			\includegraphics[width=4cm,height=4cm]{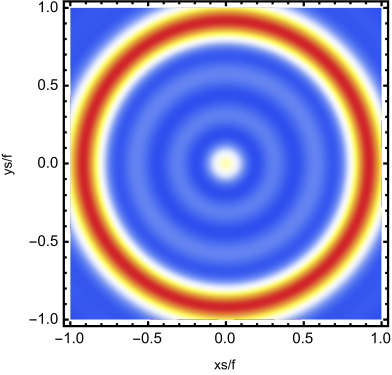}
			%\caption{fig1}
		\end{minipage}%
	}%
	\subfigure[$\omega = 50$ \label{c2}]{
		\begin{minipage}[t]{0.33\linewidth}
			\centering
			\includegraphics[width=4cm,height=4cm]{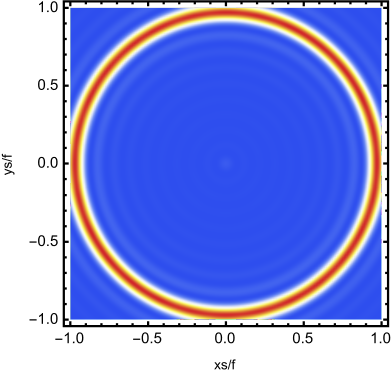}
			%\caption{fig2}
		\end{minipage}%
	}%
	\subfigure[$\omega = 80$ \label{c3}]{
		\begin{minipage}[t]{0.33\linewidth}
			\centering
			\includegraphics[width=4cm,height=4cm]{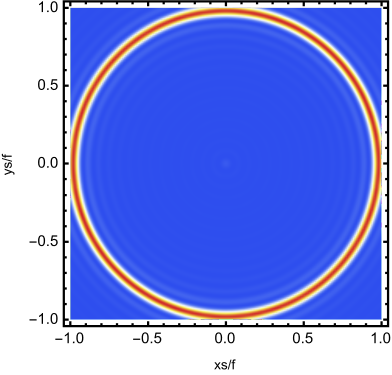}
			%\caption{fig2}
		\end{minipage}
	}%
	\quad
	\subfigure[$\omega = 20$ \label{c4}]{
		\begin{minipage}[t]{0.33\linewidth}
			\centering
			\includegraphics[width=4cm,height=4cm]{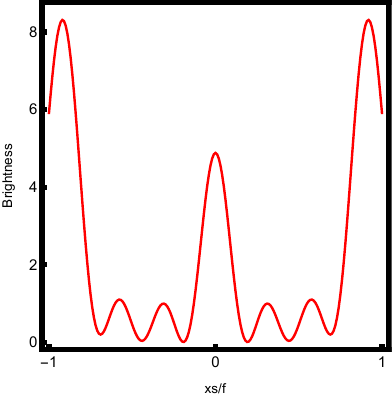}
			%\caption{fig2}
		\end{minipage}
	}%
	\subfigure[$\omega = 50$ \label{c5}]{
		\begin{minipage}[t]{0.33\linewidth}
			\centering
			\includegraphics[width=4cm,height=4cm]{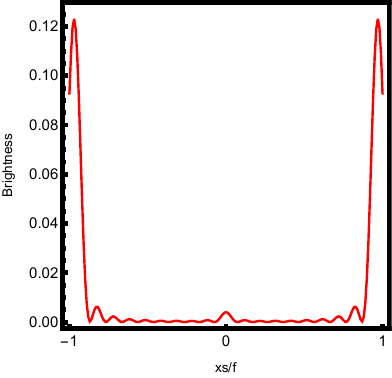}
			%\caption{fig2}
		\end{minipage}
	}%
	\subfigure[$\omega = 80$ \label{c6}]{
		\begin{minipage}[t]{0.33\linewidth}
			\centering
			\includegraphics[width=4cm,height=4cm]{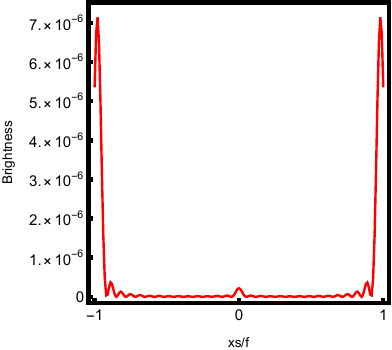}
			%\caption{fig2}
		\end{minipage}
	}%	
	\caption{Holographic images and the corresponding lensed response brightness at different source frequencies $\omega = 20, 50, 80$. In this analysis, we set $\gamma = -0.3$, $\xi_h = 1$, and $\theta_{obs} = 0$.}
	\label{c16}
\end{figure}
In Fig.~\ref{c16}, we explore the effect of varying frequencies on the holographic Einstein images generated by the wave source. Figs.~\ref{c1}, \ref{c2}, and \ref{c3} demonstrate that the rings become progressively sharper as the frequency increases. This sharpening is expected, as the geometric optics approximation becomes increasingly accurate at higher frequencies, allowing for finer image details to be captured. Additionally, Figs.~\ref{c4}, \ref{c5}, and \ref{c6} depict the corresponding lensing response function for the holographic Einstein images. The results suggest that as the frequency increases, there is a noticeable reduction in the brightness of the lensing response. This decrease in brightness may be attributed to the diminishing diffraction effects at higher frequencies, which leads to a more concentrated but less intense image. Overall, these findings highlight the critical role that frequency plays in determining the resolution and brightness of holographic Einstein images, with higher frequencies enhancing the precision of the observed features while concurrently reducing the overall intensity of the lensing response.

\begin{figure}[!ht]
	\centering
	\subfigure[$\xi_h = 4$\label{d1}]{
		\begin{minipage}[t]{0.33\linewidth}
			\centering
			\includegraphics[width=4cm,height=4cm]{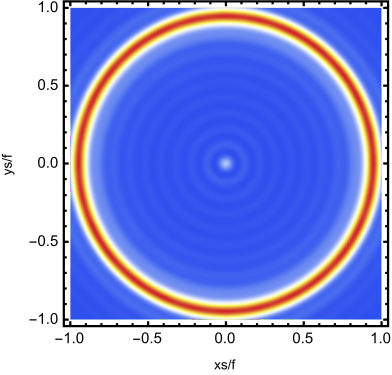}
			%\caption{fig1}
		\end{minipage}%
	}%
	\subfigure[$\xi_h = 6$\label{d2}]{
		\begin{minipage}[t]{0.33\linewidth}
			\centering
			\includegraphics[width=4cm,height=4cm]{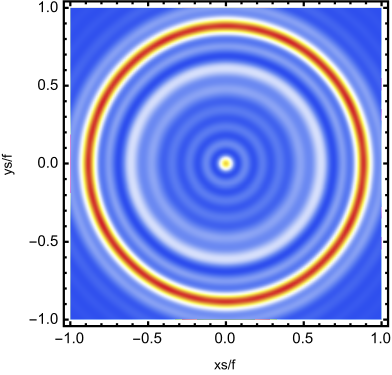}
			%\caption{fig2}
		\end{minipage}%
	}%
	\subfigure[$\xi_h = 8$\label{d3}]{
		\begin{minipage}[t]{0.33\linewidth}
			\centering
			\includegraphics[width=4cm,height=4cm]{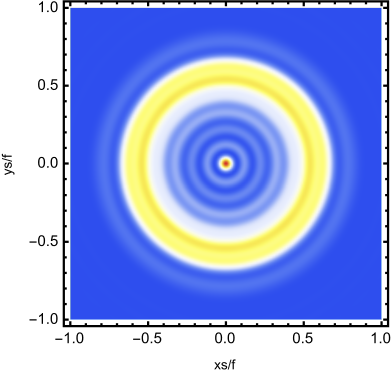}
			%\caption{fig2}
		\end{minipage}
	}%
	\quad
	\subfigure[$\xi_h = 4$\label{d4}]{
		\begin{minipage}[t]{0.33\linewidth}
			\centering
			\includegraphics[width=4cm,height=4cm]{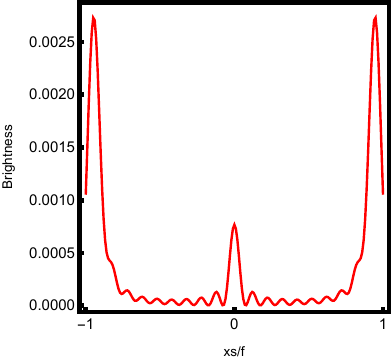}
			%\caption{fig2}
		\end{minipage}
	}%
	\subfigure[$\xi_h = 6$\label{d5}]{
		\begin{minipage}[t]{0.33\linewidth}
			\centering
			\includegraphics[width=4cm,height=4cm]{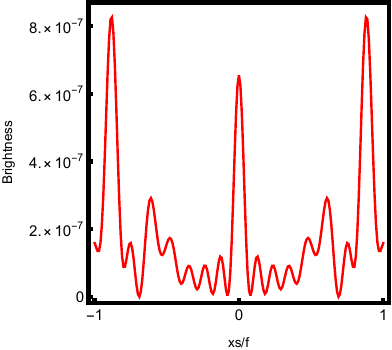}
			%\caption{fig2}
		\end{minipage}
	}%
	\subfigure[$\xi_h = 8$\label{d6}]{
		\begin{minipage}[t]{0.33\linewidth}
			\centering
			\includegraphics[width=4cm,height=4cm]{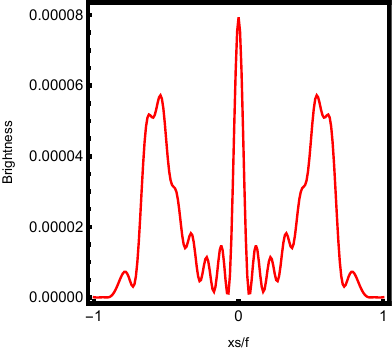}
			%\caption{fig2}
		\end{minipage}
	}%	
	\caption{Holographic images and the corresponding lensed response brightness at different temperature. In this analysis, we set $\gamma = -0.3$, $\omega = 50$, and $\theta_{obs} = 0$.}
	\label{d16}
\end{figure}
\begin{figure}[!ht]
	\centering
	\subfigure[$\xi_h = 0.2$\label{e1}]{
		\begin{minipage}[t]{0.33\linewidth}
			\centering
			\includegraphics[width=4cm,height=4cm]{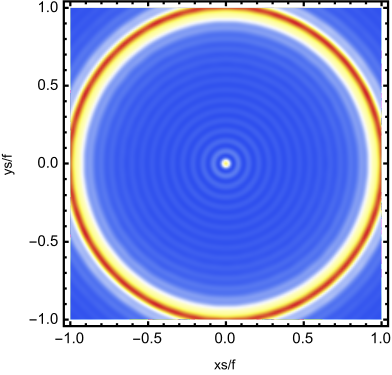}
			%\caption{fig1}
		\end{minipage}%
	}%
	\subfigure[$\xi_h = 1.2$\label{e2}]{
		\begin{minipage}[t]{0.33\linewidth}
			\centering
			\includegraphics[width=4cm,height=4cm]{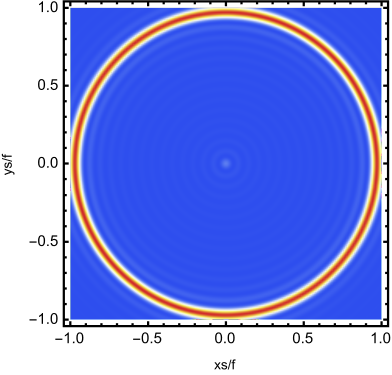}
			%\caption{fig2}
		\end{minipage}%
	}%
	\subfigure[$\xi_h = 2.2$\label{e3}]{
		\begin{minipage}[t]{0.33\linewidth}
			\centering
			\includegraphics[width=4cm,height=4cm]{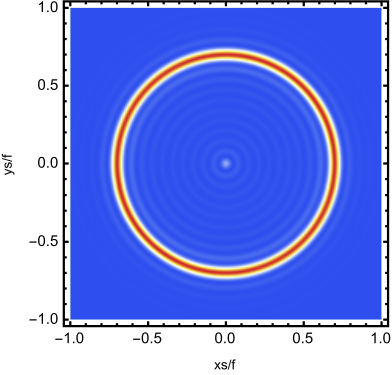}
			%\caption{fig2}
		\end{minipage}
	}%
	\quad
	\subfigure[$\xi_h = 0.2$\label{e4}]{
		\begin{minipage}[t]{0.33\linewidth}
			\centering
			\includegraphics[width=4cm,height=4cm]{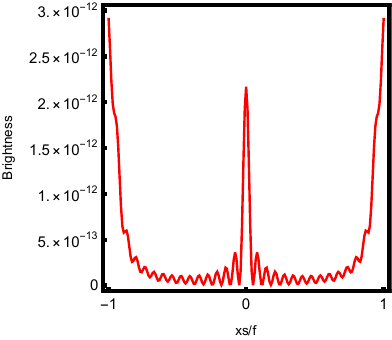}
			%\caption{fig2}
		\end{minipage}
	}%
	\subfigure[$\xi_h = 1.2$\label{e5}]{
		\begin{minipage}[t]{0.33\linewidth}
			\centering
			\includegraphics[width=4cm,height=4cm]{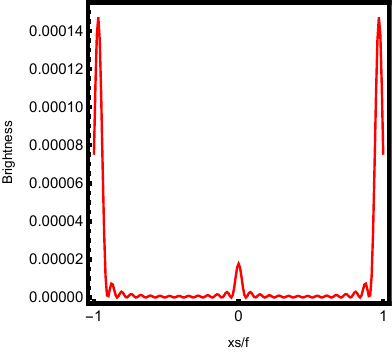}
			%\caption{fig2}
		\end{minipage}
	}%
	\subfigure[$\xi_h = 2.2$\label{e6}]{
		\begin{minipage}[t]{0.33\linewidth}
			\centering
			\includegraphics[width=4cm,height=4cm]{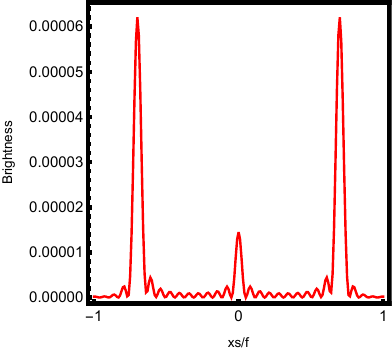}
			%\caption{fig2}
		\end{minipage}
	}%	
	\caption{Holographic images and the corresponding lensed response brightness at different temperature. In this analysis, we set $\gamma = 0.3$, $\omega = 80$, and $\theta_{obs} = 0$.}
	\label{e16}
\end{figure}
\begin{figure}[!ht]
	\centering
	{
		\begin{minipage}[t]{0.5\linewidth}
			\centering
			\includegraphics[width=6cm,height=6cm]{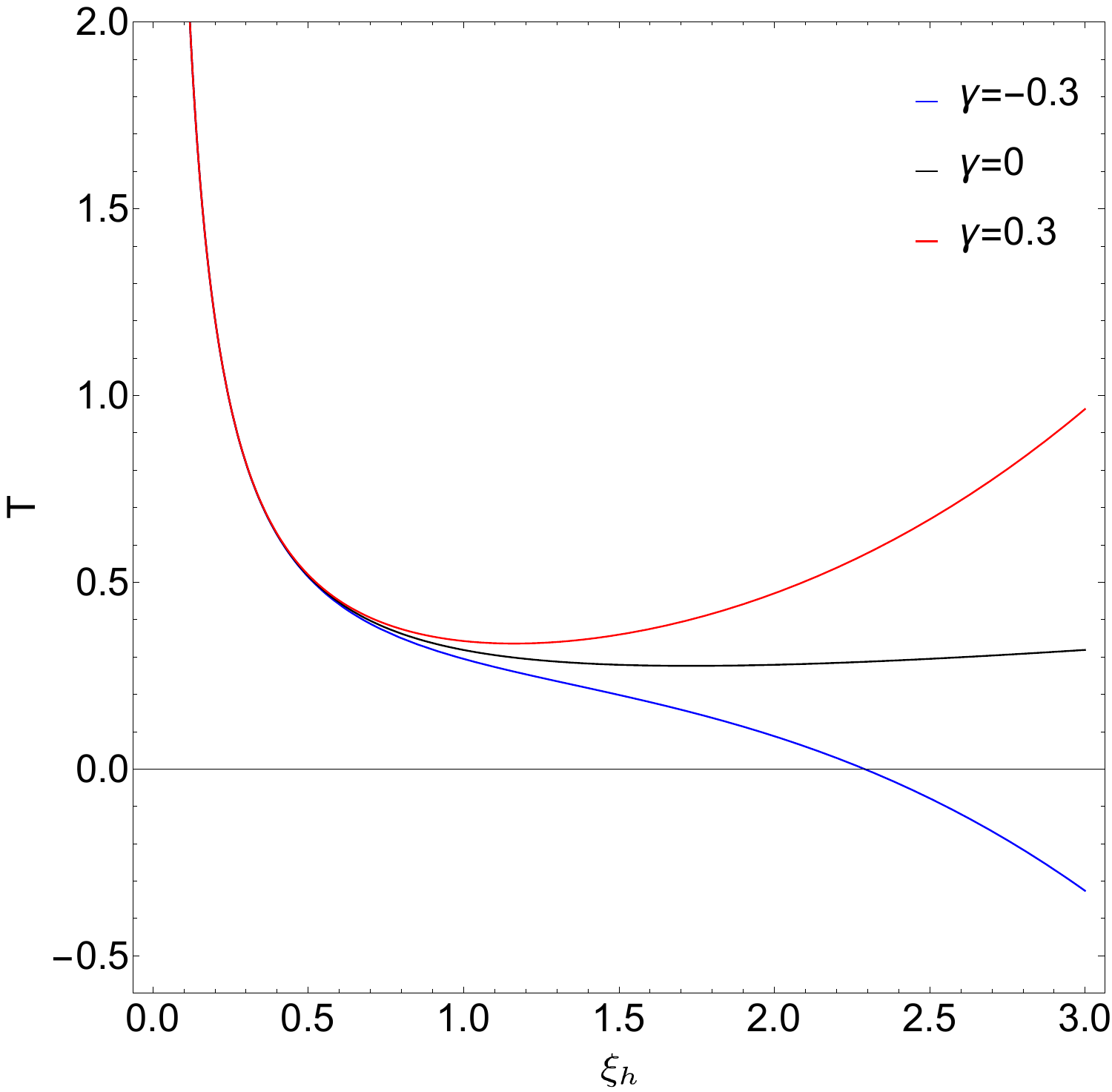}
			%\caption{The effective potential for various parameters $\gamma$}
		\end{minipage}%
	}%
	\caption{The temperature of the AdS BH in Horndeski theory is represented by three lines: the blue line corresponds to $\gamma=-0.3$, the black line corresponds to $\gamma=0$, and the red line corresponds to $\gamma=0.3$.} 
	\label{T3}
\end{figure}

Additionally, we examined the impact of horizon temperature on the holographic Einstein images for the parameter $\gamma = -0.3$, as illustrated in Fig.~\ref{d16}. In Figs.~\ref{d1} and \ref{d4}, when the horizon radius is $\xi_h = 4$ and the horizon temperature is $T = -1.14989$, the holographic Einstein image is characterized by a series of axially symmetric bright rings and a central bright spot. The brightest ring is more luminous and positioned farther from the central spot. As shown in Figs.~\ref{d2} and \ref{d5}, when the horizon radius is increased to $\xi_h = 6$ and the temperature decreases to $T = -4.63937$, the radii of these bright rings reduce, causing the brightest ring to move closer to the central spot. Upon further increasing the horizon radius to $\xi_h = 8$, as depicted in Figs.~\ref{d3} and \ref{d6}, with the horizon temperature now at $T = -11.5566$, the radii of the bright rings continue to shrink, and the brightness of the brightest ring diminishes, becoming less intense than that of the central spot. These findings suggest that as the horizon radius $\xi$ increases, leading to a decrease in horizon temperature, the radii of the bright rings progressively diminish, while the central spot becomes relatively brighter compared to the rings. This behavior underscores the intricate relationship between horizon temperature and the morphology of holographic Einstein images, indicating that changes in thermodynamic properties of the BH could be directly reflected in the observed image structure. We also examined the effect of horizon temperature on the holographic Einstein images for the parameter $\gamma = 0.3$, as illustrated in Fig.~\ref{e16}. As the radius of the event horizon increases from $\xi_h=0.2$ ($T=1.20977$) to $\xi_h=1.2$ ($T=0.33569$) and then to $\xi_h=2.2$ ($T=0.537787$), the ring gradually contracts. This result is consistent with the conclusions drawn from Fig.~\ref{d16}. However, we observe that the correlation between the ring radius and temperature is not very pronounced. This is because, when the parameter $\gamma$ in Horndeski theory is positive, the BH temperature is no longer a monotonic function of the event horizon, as illustrated in Fig.~\ref{T3}.
\textbf{}

Analyzing the holographic Einstein ring solely through the lens of wave optics provides an incomplete perspective on the system's behavior. To achieve a more comprehensive understanding, we now approach the relationship between the photon ring and the holographic Einstein ring from the standpoint of geometric optics. In describing photon motion, two conserved quantities play a crucial role: the energy of the photon, $E = f(r) (\partial t/\partial \lambda)$, and its angular momentum, $\tilde{L} = r^2 (\partial \varphi/\partial \lambda)$, where $\lambda$ is the affine parameter. Considering photon motion in the equatorial plane $(\theta = \pi / 2)$, its four-velocity vector $v^\alpha = \left(d/d \lambda\right)^\alpha$ satisfies the following condition:
\begin{equation}
	-f(r)\left(\frac{d t}{d \lambda}\right)^2 + f(r)^{-1}\left(\frac{d r}{d \lambda}\right)^2 + r^2 \sin^2 \theta \left(\frac{d \varphi}{d \lambda}\right)^2 = 0,
\end{equation}
which simplifies to the equation
\begin{equation}
	\dot{r}^2 = E - z(r) \tilde{L} ,
\end{equation}
where
\begin{equation}
	z(r) = f(r)/r^2.
\end{equation}

Building on Refs.~\cite{Liu:2022cev,Zeng:2023zlf,Zeng:2023tjb}, we consider an incident angle defined by the boundary's normal vector $\theta_{in}$ and the direction $n^\zeta = \partial/\partial r^\zeta$. Under these conditions, we obtain
\begin{equation}
	\cos \theta_{i n}=\left.\frac{g_{\alpha \beta} v^\alpha n^\beta}{|v||n|}\right|_{r=\infty}=\left.\sqrt{\frac{\dot{r}^2 / f}{\dot{r}^2 / f+\tilde{L} / r^2}}\right|_{r=\infty}.
\end{equation}
Therefore, we derive
\begin{equation}
	\sin^2 \theta_{\text{in}} = 1 - \cos^2 \theta_{\text{in}} = \left.\frac{\tilde{L}^2 z(r)}{\dot{r}^2 + \tilde{L}^2 z(r)}\right|_{r \to \infty} = \frac{\tilde{L}^2}{E^2}.
\end{equation}
This relationship indicates that the incident angle $\theta_{\text{in}}$ is connected to the photon's energy $E$ and angular momentum $\tilde{L}$ as
\begin{equation}
	\sin \theta_{\text{in}} = \frac{\tilde{L}}{E}.
\end{equation}

\begin{figure}[!ht]
	\centering
	\subfigure[\label{qq2}]{
		\begin{minipage}[t]{0.5\linewidth}
			\centering
			\includegraphics[width=8cm,height=4.8cm]{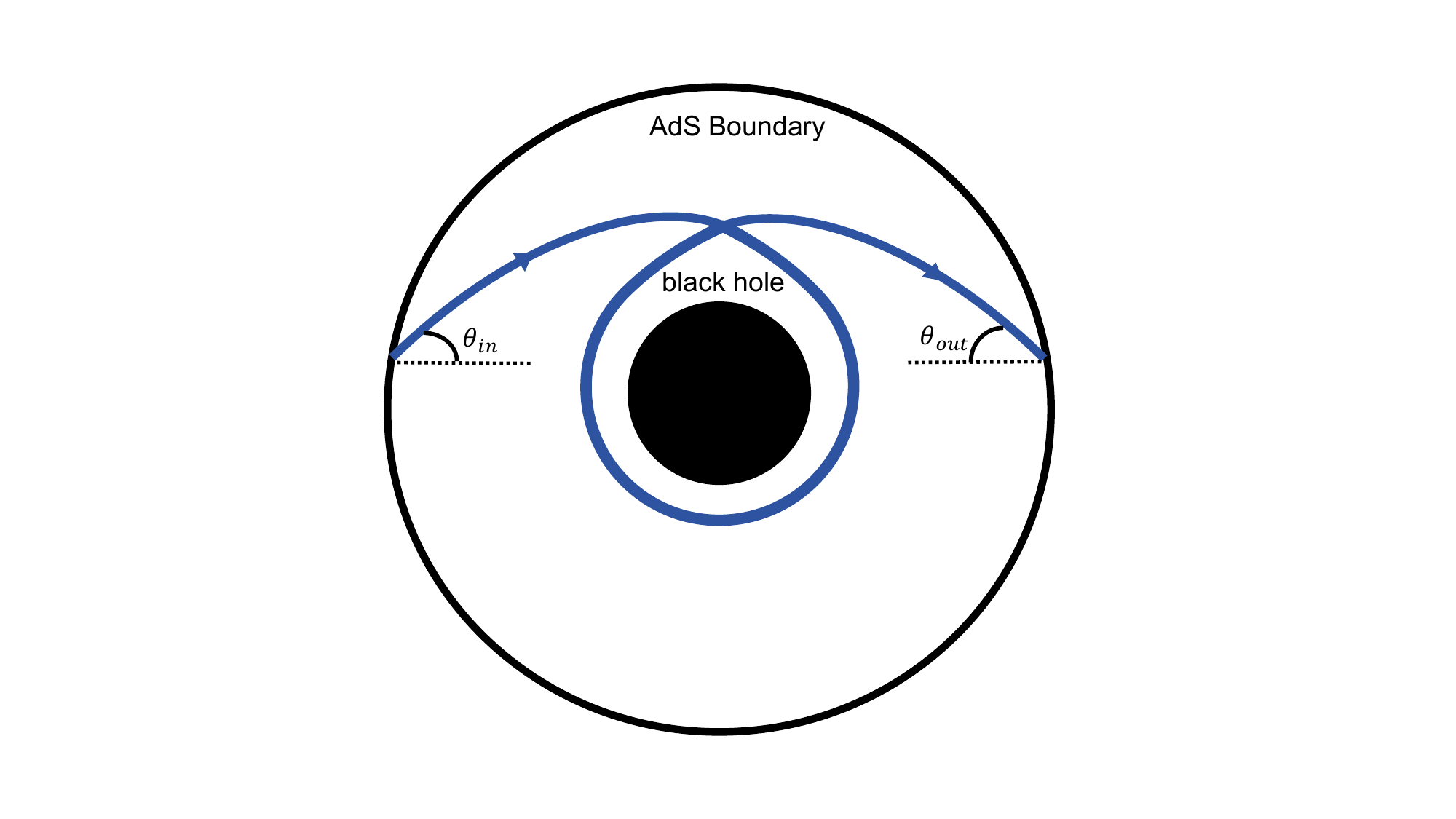}
			%\caption{The effective potential for various parameters $\gamma$}
		\end{minipage}%
	}%
	\subfigure[\label{qq3}]{
		\begin{minipage}[t]{0.5\linewidth}
			\centering
			\includegraphics[width=8cm,height=4.8cm]{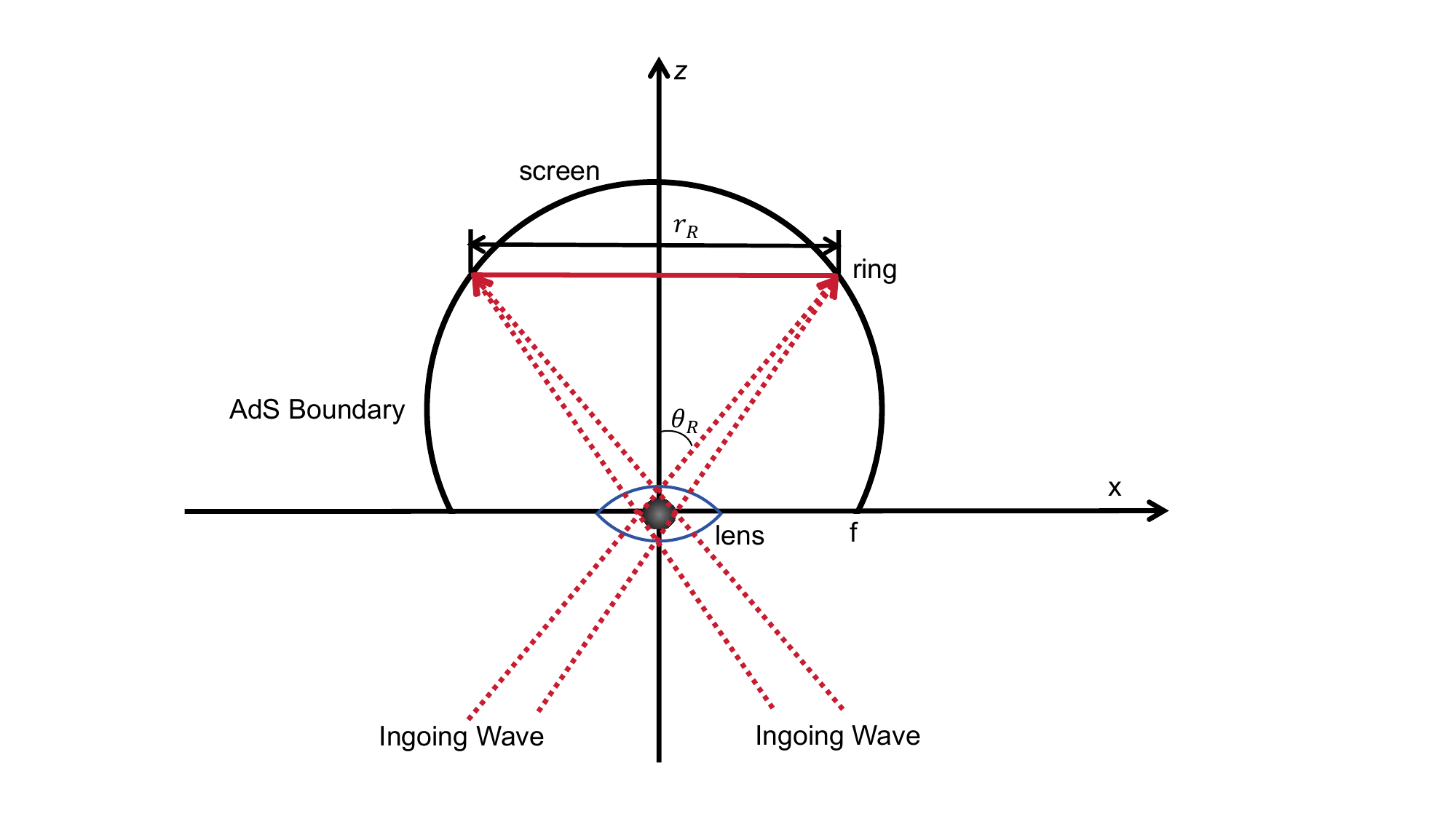}
			%\caption{The effective potential for various parameters $\gamma$}
		\end{minipage}%
	}%
	\caption{(a). The trajectory of an incident photon completes a full rotation around the BH. (b). The relationship between the ring radius and the ring angular.} 
	\label{figures11}
\end{figure}
\begin{figure}[!ht]
	\centering
	\subfigure[$\gamma=-0.3$\label{point1}]{
		\begin{minipage}[t]{0.35\linewidth}
			\centering
			\includegraphics[width=4cm,height=4cm]{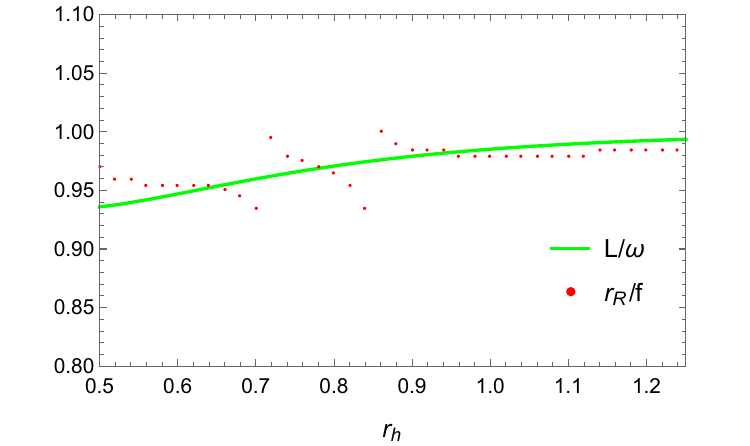}
			%\caption{fig1}
		\end{minipage}%
	}%
	\subfigure[$\gamma=0$\label{point2}]{
		\begin{minipage}[t]{0.35\linewidth}
			\centering
			\includegraphics[width=4cm,height=4cm]{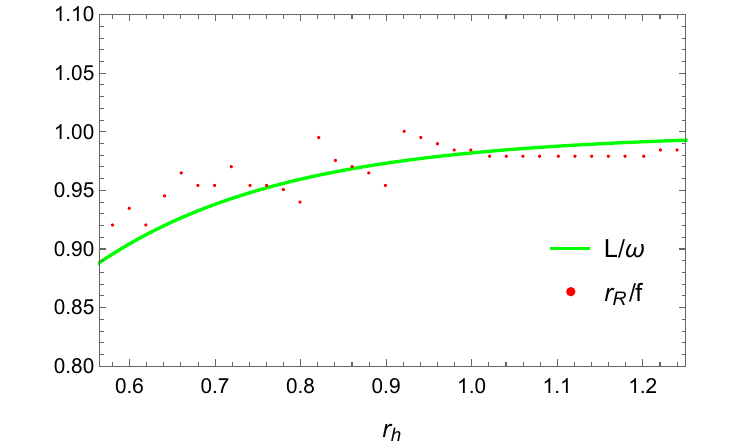}
			%\caption{fig2}
		\end{minipage}%
	}%
	\subfigure[$\gamma=0.3$\label{point3}]{
		\begin{minipage}[t]{0.35\linewidth}
			\centering
			\includegraphics[width=4cm,height=4cm]{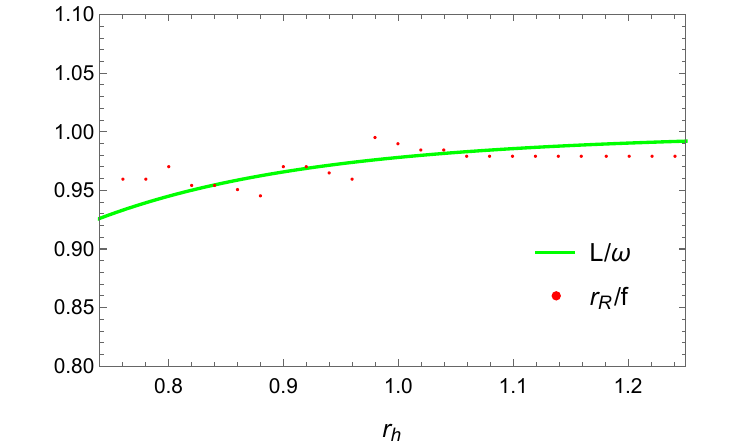}
			%\caption{fig2}
		\end{minipage}
	}%
	\caption{The Einstein ring radius is shown as a function of the event horizon under different values of the parameter $\gamma$.
	}
	\label{point333}
\end{figure}

\textbf{}
As shown in Fig.~\ref{qq2}, when the photon's entry and exit points align with the BH's center along a straight line, its trajectory forms a ring-shaped image. Upon reaching the photon ring, the photon enters a critical orbit around the BH. We denote the angular momentum of the photon in this critical state as $L$, and its orbital equation satisfies
\begin{equation}
	\dot{r}=0, \frac{{~d} z }{{d} r}=0.
\end{equation}
We define the radius of the circle in Fig.~\ref{qq3} as $r_R$, leading to the relation
\begin{equation}
	\sin \theta_R = \frac{r_R}{f}.
\end{equation}
For sufficiently large angular momentum, we have $\sin \theta_R = \sin \theta_{\text{in}}$, and thus, we derive the relationship
\begin{equation}\label{eqrrf}
	\frac{r_R}{f} = \frac{L}{E}.
\end{equation}
To validate Eq.~(\ref{eqrrf}), we present in Fig.~\ref{point333} the Einstein ring radius under different event horizons. The results demonstrate that the incident angle obtained via geometric optics agrees well with the observational angle derived from wave optics across various parameters $\gamma$. Moreover, this conclusion holds regardless of the parameter $\gamma$ in Horndeski theory.

\section{Conclusions}
\label{Concl}
In this work, we have investigated the holographic Einstein rings in the context of AdS BHs within Horndeski theory, employing both wave optics and geometric optics methods. Our analysis reveals that the characteristics of the holographic Einstein ring, such as its radius and brightness, are influenced by multiple factors, including the observer’s position, the properties of the wave source, the BH’s event horizon, temperature, and the parameter $\gamma$ in Horndeski theory.

We have demonstrated that when the observer is located at the north pole of the AdS boundary and the wave source is placed at the south pole, the resultant image forms a circular Einstein ring with a central Poisson spot, coinciding with the location of the BH's photon sphere. The size and brightness of this ring are affected by the wave source’s frequency and the BH’s temperature. As the source frequency increases, the ring sharpens, while the overall brightness decreases due to reduced diffraction effects. On the other hand, as the BH's temperature increases, the radius of the event horizon and the Einstein ring both shrink, causing the central Poisson spot to become more prominent. Furthermore, we examined the relationship between the Einstein ring and BH parameters, including the parameter $\gamma$ in Horndeski theory. Our findings indicate that the ring’s morphology and brightness are significantly influenced by the parameter $\gamma$. As $\gamma$ increases, the Einstein ring evolves from symmetric concentric rings into arcs or even distinct bright spots depending on the observer's angle. 

In addition, we explored the effect of horizon temperature on the Einstein ring structure. As the event horizon radius increases, resulting in lower temperatures, the ring’s radius shrinks, and the central Poisson spot becomes more prominent. This behavior underscores the sensitive relationship between the thermodynamic properties of the BH and the observed holographic image. The results also indicate that the influence of the BH's temperature varies depending on the value of the parameter $\gamma$. For positive $\gamma$, the BH temperature is no longer a monotonic function of the event horizon, leading to more complex ring behaviors. The comparison between wave optics and geometric optics further verifies that the observed angle in wave optics corresponds well with the incident angle in geometric optics, validating the consistency between these two approaches for analyzing the Einstein ring.

Overall, this study provides a deeper understanding of the holographic images of BHs in Horndeski theory and highlights the potential of using holographic Einstein rings as a tool to probe the underlying geometry and thermodynamic properties of BHs. Future research can extend this approach to other gravitational theories and BH spacetimes, further enriching our understanding of holographic dualities and BH physics.

\vspace{10pt}

\noindent {\bf Acknowledgments}

\end{document}